\documentclass[10pt, leqno]{amsart}

\usepackage[
    top=2.5cm,
    bottom=2.5cm,
    left=2.5cm,
    right=2.5cm,
    heightrounded,
]{geometry} % More reliable way to set margins
\usepackage{csquotes} % For linguistic quotes, good practice
\usepackage{amsmath}
\usepackage{amsthm}
\usepackage{amssymb}

\usepackage{graphicx}
\usepackage{hyperref}
\usepackage{xcolor}
\usepackage[T1]{fontenc} % For better font encoding and character support
\usepackage{titlesec}
\titleformat{\section}{\normalfont\Large\bfseries}{\thesection}{1em}{}
\hypersetup{
    colorlinks=true,
    linkcolor=black,
    citecolor=black,
    filecolor=black,
    urlcolor=black,
}

\numberwithin{equation}{section}
\newtheorem{theorem}{Theorem}[section] % Numbered within sections (e.g., Theorem 2.1)
\newtheorem{definition}[theorem]{Definition}

\renewcommand{\L}{\mathcal{L}}
\newcommand{\R}{\mathbb{R}} % Correctly define \R for Real Numbers
\title[Conformal Solitons on Vaidya Spacetime]{Solutions and Gradient of the Conformal Ricci Bourguignon Soliton on Vaidya Spacetime}

\author[Ayaan AR]{Ayaan Abdur Rehman \textsuperscript{1}}
\address{Dept. of Physics and Astronomy, National Institute of Technology, Rourkela, Odisha, India}
\email{ayaanar1908@gmail.com \textsuperscript{(1)}}

\author[N. S. Iyer]{Narayan S Iyer \textsuperscript{2}}
\address{Dept. of Physics and Astronomy, National Institute of Technology, Rourkela, Odisha, India}
\email{narayansiyer7@gmail.com \textsuperscript{(2)}}

\author[N. A. Pundeer]{Naeem Ahmed Pundeer \textsuperscript{3}}
\address{Dept. of Mathematics, Jadavpur University, Kolkata, West Bengal, India}
\email{pundir.naeem@gmail.com \textsuperscript{(3)}}

\date{\today}

\begin{document}

\maketitle

\begin{abstract}
In this work, we derive the complete and explicit solution for the conformal Ricci-Bourguignon soliton on Vaidya spacetime. We provide the closed-form expression for the vector field and establish the necessary conditions for the existence of the scalar potential, for which we also derive an explicit form. Our solution to the underlying system of linear partial differential equations proves that such solitons exist if and only if the mass function vanishes, forcing the metric to reduce to flat Minkowski spacetime (Schwarzschild, $m=0$). Synthesizing prior works, we show that the established classification of the soliton as shrinking, steady, or expanding is justified by the principles of linear stability. These findings refine the set of possible solitons within the non-linear theory of geometric flows by proving they are only admissible in the non-radiating vacuum limit, thereby enhancing the reliability of such models.
\end{abstract}

% --- TABLE OF CONTENTS ---
% This will appear on a new page after the abstract.
\newpage
\tableofcontents
\newpage
\section{Preliminaries and Introduction} 
\vspace{0.25cm}
The Ricci Bourguignon flow is a generalization of Ricci flow on a manifold that was first suggested as a field of study by J.P Bourguignon in \cite{10.1007/BFb0088841} . This was done on the basis of papers by Lichnerowicz and Aubin \cite{aubin1970metriques} . It was then coined as the Ricci-Bourguignon flow in \cite{catino2016gradient},\cite{catino2017ricci}.
As seen in (\cite{catino2016gradient},\cite{catino2017ricci}) we go with the following definition
\vspace{0.5cm}
\begin{definition}On a Riemannian manifold (M,g) the Ricci Bourguignon flow is 
$$\partial_t g=-2S+2\rho Rg $$
Where $\rho \in \R$, S is the Ricci tensor, g is the metric tensor    
\end{definition}
\vspace{0.5cm} 

As demonstrated by \cite{catino2017ricci} and observation, the right-hand side of the above equation reduces to 
\begin{align}
 \rho=0 \implies \partial_t g &=-2S 
\\
  \rho=\frac{1}{2} \implies  \partial_t g &=-2S + Rg 
\\
  \rho=\frac{1}{n}  \implies \partial_t g &= -2S + \frac{2Rg}{n}
\end{align}  
Note that the right handside of the above equations are Ricci tensor, Einstein tensor, traceless Ricci Tensor respectively and are obtained by changing the value of $\rho$ in the ricci bourguignon flow. 
\vspace{1cm}

In 2005 A.E Fischer in his paper \cite{fischer2004introduction} formulated the conformal ricci flow. Fischer called these flows conformal due to the involvement of conformal geometry.In conformal ricci flow, we replace the usual volume constraint in ricci flow by a curvature constraint I.e 

$$\{vol(M,g_t)=1\} \rightarrow \{R(g_t)=-1\} $$

\vspace{0.25cm}
For a more detailed explanation, see Ref. fischer in \cite{fischer2004introduction} \\
Therefore in summary based on fischer's work we get the next definition
\vspace{0.5cm}
\begin{definition} given a riemannian manifold (M,g) and $DIM(M)\geq3$ the conformal ricci flow is defined as
$$\partial_t g =- \Bigl(p+\frac{1}{n}\Bigl)g - 2S $$
$$R(g)=-1 $$
    Where $n=DIM(M)$, p is conformal pressure,R(g) is the curvature scalar.
\end{definition}

The reason p is called conformal pressure is due to its similarity to a similar analogy to the navier-stokes equation \cite{fischer2004introduction}.On the basis of this, Shubham dwivedi constructed the ricci bourguignon solitons \cite{dwivedi2021some} , \cite{alhouiti2024geometric}.
based on this we reach the third definition
\vspace{0.5cm}

\begin{definition} A Vector field X on a given a riemannian manifold (M,g) satisfies the Ricci Bourguignon soliton if
$$\L_X g +2S=2\bigl( \beta g + \alpha Rg \bigl)$$
where S is the Ricci Tensor and $\alpha $ , $\beta $ are constants     
\end{definition}
\vspace{0.25cm}

Solitons generate self similar solutions to the corresponding flow on the manifold \cite{catino2017ricci}
. Recently in the year 2024, on the basis of the paper \cite{basu2015conformal}by basu and ariesh bhattachariya on the formulation of conformal ricci soliton  , \cite{alhouiti2024geometric}      formulated the conformal ricci bourguignon soliton as follows 
\vspace{0.5cm}

\begin{definition}\label{def:6} A vector field X on a riemannian manifold (M,g) is a conformal ricci bourguignon soliton if 
$$\L_X g +2S=\Bigl(2\beta-\Big(p+\frac{2}{n}\Big)\Big)g +2\alpha Rg $$
where p is a time invariant scalar field ( i.e constant as a function of time ).     
\end{definition}

\newpage
\section{An aside on Vaidya Spacetime}  
\vspace{0.25cm}
The Vaidya Metric was constructed by C. Vaidya in his paper \cite{chunilal1943external}. The metric was obtained by modifying the schwartzchild metric by making the mass a function of the null Co ordinate namely 'u' .
\newline
Vaidya observed the symmetry in his equation and used a coordinate system that we now call the Eddington finkelstein coordinates. This gave the metric the following form in terms of differentials \cite{vaidya1953newtonian} \cite{griffiths2009exact}
$$ds^2=\biggl(\frac{2m}{r} - 1\biggl)du^2 - 2drdu+r^2d\theta^2 +r^2\sin^2 \theta d\phi^2 $$
Where $m \equiv m(u) $
\vspace{0.5cm}
The Eddington finkelstein coordinate transformation converts the (t, x, y, z) Co ordinates into (u, r, $\theta$, $\phi $). The transformation is identical to that of spherical coordinates except for u (which is also called the null Co ordinate). The transformation is for u is as follows:
       $$ u=t-(r+2m \ln(r-2m)) $$     

See \cite{griffiths2009exact} for further details of the conversion to Eddington Finkelstein Co ordinates.
 
\vspace{0.2cm}

Upon converting the metric tensor into its matrix representation we end up with the following definition:-
\begin{definition} The vaidya metric as a matrix in Eddington Finkelstein coordinates is defined as 
 $$g_{\mu\nu} =\begin{bmatrix}
  \frac{2m-r}{r} & - 1 & 0 & 0 \\
  - 1            &   0 & 0 & 0 \\
    0            &   0 & r^2 & 0\\
    0            &  0  & 0  & r^2\sin^2\theta
  
 \end{bmatrix}$$
\end{definition}
Now we shall proceed by listing down the non zero components of the ricci curvature tensor in the Eddington finkelstein coordinates as seen in the following papers

 The non zero components Riemann curvature tensor are the following (see ref \cite{shaikh2017curvature})  \begin{align*}
R_{1212} &= \frac{-2m}{r^3}, \\
R_{1313} &= \frac{-2m + r^2m' - m}{r^2}, \\
R_{1323} &= \frac{m}{r}, \\
R_{1424} &= \frac{m \sin^2\theta}{r}, \\
R_{1414} &= \frac{-\bigl(2m^2 - mr + r^2m'\bigr)\sin^2\theta}{r^2}, \\
R_{3434} &= 2mr\sin^2\theta.
\end{align*}

We can further obtain the expression of the ricci tensor by contracting the riemann curvature in the following format $$Ric_{\mu\nu}=  R^a_{ a\mu\nu} $$
for our particular case we end up with the following value of Ricci tensor
$$Ric=\frac{2\dot m(u) }{r^2} du\otimes du$$
See \cite{shaikh2017curvature} for more information.
\\ 
For further applications sake, we shall go with the following representation of the Ricci curvature tensor.
\begin{definition}The ricci curvature tensor(Ric or S) of vaidya space time is the following when written in matrix format
$$Ric_{\mu\nu}=\begin{bmatrix}
 \frac{2\dot m(u)}{r^2} & 0&0&0\\
 0&0&0&0\\
 0&0&0&0\\
 0&0&0&0
\end{bmatrix}$$

\end{definition}
\vspace{0.8cm}
 Note that the Vaidya metric is an Einstein metric (i.e., it's Ricci curvature scalar is zero ,see \cite{10.1007/BFb0088841} section 1.6). This can also be manually verified upon using the definition of the Ricci Scalar by raising the index of the ricci tensor and taking its trace (i.e) 
 $$R=g^{\mu\nu}Ric_{\mu\nu}=0 $$
 Where $g^{\mu\nu}$ is the inverse metric (which can obtained from $g_{\mu\lambda}g^{\lambda\nu}= \delta_{\mu}^{\nu} $) which is written following in matrix format
 $$g^{\mu\nu}=\begin{bmatrix}
   0&-1&0&0\\
   -1& \bigl(1-\frac{2m}{r}\bigl)&0&0\\
   0&0&r^2&0\\
   0&0&0&r^2\sin^2\theta
 \end{bmatrix}$$
 A point to be noted is that the Eddington finkelstein transformation is $C^\infty$ (except at the singularity at m) and hence the transformed Vaidya metric remains an einstein metric(I.e trace free ) even after coordinate transformation(see remark 3.22 in \cite{10.1007/BFb0088841} ) 
 
\vspace{0.1cm}
\begin{definition}
    Given a Metric Tensor Field ( say g ), we define the lie derivitive of the tensor field with respect to a vector field (abbreviated as $\L_{X} g$ ) and \\
    Where $X=X^\alpha \partial_\alpha$ as
 \begin{align*}
    \L_X g = (\L_X g)_{ij}dx^i\otimes dx^j \\
    (\L_X g) _{ij} = (\L_X g) (\partial _i, \partial_j)   \\ 
     = X^k\partial _k(g_{ij}) + g_{kj}(\partial _i X^k) +g_{ik}(\partial _j X^k) \\
\end{align*}
\end{definition} 

    We shall hence proceed to compute the components of the lie derivitive.     
   
\begin{align}
    (\L_X g)_{11} &= 2\left(\frac{2m - r}{r}\right) \partial_u A - \partial_u B, \label{2.1} \\
    (\L_X g)_{12} &= \frac{2m - r}{r}\partial_r A - \partial_r B - \partial_u A,  \label{2.2} \\
    (\L_X g)_{13} &= \frac{2m - r}{r}\partial_\theta A - \partial_\theta B + r^2 \partial_u C,  \label{2.3} \\
    (\L_X g)_{14} &= \frac{2m - r}{r}\partial_\phi A - \partial_\phi B + r^2 \sin^2\theta \partial_u D,  \label{2.4} \\
    (\L_X g)_{21} &= (\L_X g)_{12},  \label{2.5} \\
    (\L_X g)_{22} &= -2\partial_r A,  \label{2.6} \\
    (\L_X g)_{23} &= r^2 \partial_r C - \partial_\theta A, \label{2.7} \\
    (\L_X g)_{24} &= r^2 \sin^2\theta \partial_r D - \partial_\phi A,  \label{2.8} \\
    (\L_X g)_{31} &= (\L_X g)_{13}, \label{2.9} \\
    (\L_X g)_{32} &= (\L_X g)_{23},  \label{2.10} \\
    (\L_X g)_{33} &= 2(rB + r^2 \partial_\theta C),  \label{2.11} \\
    (\L_X g)_{34} &= r^2 \partial_\phi C + r^2 \sin^2\theta \partial_\theta D,  \label{2.12} \\
    (\L_X g)_{41} &= (\L_X g)_{14},  \label{2.13} \\
    (\L_X g)_{42} &= (\L_X g)_{24},  \label{2.14} \\
    (\L_X g)_{43} &= (\L_X g)_{34}, \label{2.15} \\
    (\L_X g)_{44} &= 2rB \sin^2\theta + 2r^2 C \sin\theta \cos\theta + 2r \partial_\phi D. \label{2.16}
\end{align}

\vspace{0.5cm}

\newpage
\section{The Conformal Ricci Bourguignon Soliton with respect to Vaidya Metric} 
\vspace{0.25cm}
We shall utilize [\ref{def:6}] and [\eqref{2.1}-\eqref{2.16}] we obtain the following differential equations. 
\begin{align}
\frac{2m-r}{r}\partial_uA - \partial_uB + \frac{m'A}{r} - \frac{mB}{r^2}+\frac{2m' }{r^2 }&=\frac{1}{2}\kappa\frac{2m-r}{r} \label{eqn1}\\
\partial_r A &=0 \label{eqn2} \\
rB+r^2\partial_\theta C&=\frac{\kappa r^2} {2} \label{eqn3} \\
\sin^2\theta B+ r\sin\theta\cos\theta C +r\sin^2\theta \partial_\phi D &= \frac{r\kappa\sin^2\theta}{2} \label{eqn4} \\
\frac{2m-r}{r}\partial_rA - \partial_rB-\partial_uA&=-\kappa     \label{eqn5} \\
\frac{2m-r}{r}\partial_\theta A - \partial_\theta B +r^2\partial_u C&=0 \label{eqn6} \\
\frac{2m-r}{r} \partial_\phi A - \partial_\phi B +r^2\sin^2\theta\partial_uD &= 0 \label{eqn7} \\
r^2\partial_rC - \partial_\theta A&=0 \label{eqn8} \\
r^2\sin^2\theta\partial_rD - \partial_\phi A&=0 \label{eqn9} \\
r^2\partial_\phi C+r^2\sin^2\theta\partial_\theta D&=0 \label{eqn10}.
\end{align} 
Here note that $\kappa$ stands for $2\beta - (p+\frac{1}{2}) $
\vspace{1cm}
\\
Note that Vaidya Spacetime has a singularity at r=0, this can be observed directly on the Vaidya Metric. Our manifold hence excludes the point r=0. Also observe that the Vaidya metric is Radially symmetric and hence the closed form solution also must be radially symmetric. Therefore, the closed form solution would extend to also account for division by $\sin\theta$ or $\cos\theta$.\\
From equation \eqref{eqn2} we get\\
\begin{equation*}
\partial_rA=0 \iff A=A(u, \theta, \phi) \tag{I} \label{sub:1}\\
\end{equation*}
\\
Hence both A and $\partial_uA$ are independent of r.
From equations \eqref{eqn2}, \eqref{eqn5}\\
\\
\(\ \partial_rB=\kappa-\partial_uA \)\\
\\
$\kappa$ and $\partial_uA$ are independent of R, hence\\ 
\\
\begin{equation*}
 B=(\kappa - \partial_uA)r + Q_b(u,\theta,\phi) \tag{II} \label{sub:2}
\end{equation*}
\\
from equation\eqref{eqn8}\\
\\
\(\ \partial_rC=\frac{\partial_\theta A} {r^2} \iff C=\frac{-\partial_\theta A}{r} +Q_c(u, \theta, \phi) \)\\
\\
\begin{equation*}
    C=\frac{-\partial_\theta A}{r} +Q_c(u, \theta, \phi) \tag{III} \label{sub:3}
\end{equation*}
from equation\eqref{eqn10}\\
\\
\(\ \partial_r D =\frac{\partial_\phi A} {r^2\sin^2\theta} \iff D=-\frac{\partial_\phi A} {r\sin^2\theta} +Q_d(u, \theta,\phi ) \) \\
\\
\begin{equation*}
   D=-\frac{\partial_\phi A} {r\sin^2\theta} +Q_d(u, \theta,\phi )  \tag{IV} \label{sub:4}
\end{equation*}
\\
Now, from eqn \eqref{eqn3}\\
\\
$rB+r^2\partial_\theta C=\frac{kr^2}{2} \implies B+r\partial_\theta C=\frac{kr}{2} $ \\
\\
Apply $\partial_r$ on both sides and substituting \eqref{sub:2}  we obtain\\
\\
\( (\kappa-\partial_uA)+\partial_\theta C +r\partial_{\theta r}C =\frac{\kappa}{2} \) \\
\\
Substitute equation \eqref{eqn8} and \eqref{sub:3} in above equation we get \\
\\
\(\ \frac{k}{2}-\partial_u A +\frac{\partial^2_\theta A}{r}+\partial_\theta Q_c-\frac{\partial^2_\theta A}{r} =0 \)\\
\\
Now once again from equation \eqref{eqn3} we substitute \eqref{sub:2},\eqref{sub:3}\\
\\
\(\ r(\frac{\kappa} {2}-\partial_u A )+Q_b - \frac{r}{r}\partial^2_\theta A +r\partial_\theta Q_c=0 \iff r(\partial_\theta Q_c -\partial_uA+\frac{\kappa} {2}-\frac{\partial^2_\theta A} {r} ) +Q_c =0 \)\\
\\
Upon substituting in the above equation we get\\
\\
\begin{equation*} 
 Q_b=\partial^2_\theta A \tag{V} \label{sub:5}
\end{equation*}
\\
from \eqref{eqn6} and eqn \eqref{sub:2}\\
\\
\(\ \frac{2m-r}{r} \partial_\theta A - \partial_\theta B +r^2\partial_u C=0 \)\\
\\
\(\ \implies \frac{2m-r}{r}\partial_\theta A - \partial_\theta(r(k-\partial_u A)+Q_b ) +r^2\partial_u C=0 \)\\
\\
\(\ \implies\frac{2m-r}{r}\partial_\theta A +r\partial_{u\theta} A - \partial_\theta^3+r^2\partial_u(-\frac{\partial_\theta C}{r}+Q_c)=0 \)\\
\\
\(\ \frac{2m-r} {r} \partial_\theta A - \partial_\theta^3 A +r^2\partial_uQ_c=0 \)\\
\\
Differentiate both sides by $\partial_r$. Note that $\partial_rA=0$ and $\partial_u Q_c=0$\\
\\
\(\  \implies -\frac{2m} {r^2}\partial_\theta A +2r\partial_uQ_c=0 \)\\
\\
differentiate both sides by $\partial_r^2$ we obtain \\
\\
\(\ \frac{6m}{r^3}\partial_\theta A=0 \implies \partial_\theta A=0  \)  \\
\\
Hence we obtain 
\begin{equation*}
    \partial_\theta A=0
    \tag{VI} \label{sub:6}
 \end{equation*}   \\
 
A technical detail has been with regard to the act of assuming that $\partial_\theta A =0 \text{ and/or  } m=0$ has been omitted. Here we proceed with $\partial_\theta A=0$ and 
in the Appendix we shall prove that even assuming that m=0 right at the start shall also yield \eqref{sub:6}.
\\

Note that from the previous equations(\eqref{sub:5},\eqref{sub:6}) we obtain\\
\\
\(\ \implies Q_b=0\)\\
\\
\\
\\
from \eqref{eqn7} substitute the value of B and D from equations(\eqref{sub:2},\eqref{eqn4})\\
\\
\(\ \frac{2m-r} {r}\partial_\phi A - \partial_\phi B +r^2\sin^2\theta \partial_u D=0 \)\\
\\
\(\ \frac{2m-r} {r}\partial_\phi A - \partial_\phi(r(k-\partial_u A) )+r^2\sin^2\theta\partial_u(\frac{-\partial_\phi A}{r\sin^2\theta }+Q_d)=0 \)\\
\\
differentiate both sides by $\partial_r^3$ we obtain\\
\\
\begin{equation*}
\partial_\phi A=0 \tag{VII} \label{sub:7}
\end{equation*}\\
\\
By  substituting equation(\eqref{sub:7} and \eqref{sub:2}) in \eqref{eqn7} we obtain\\
\\
\(\ r^2\sin^2\theta\partial_u D=0 \)\\
and substitute the above obtained result in \eqref{eqn4}.we hence obtain the following result
\\
\begin{equation*}
  \implies \partial_uD=\partial_u Q_d=0 \tag{VIII} \label{sub:8}
 \end{equation*}
\\
\\
From \eqref{eqn6} , substitute equations(\eqref{sub:2},\eqref{sub:3})\\ 
\\
\(\ -\partial_\theta B + r^2\partial_uC=0\)\\
\\ 
\(\ \implies -\partial_\theta(r(k-\partial_u A))+r^2\partial_uQ_c \)\\
\\
\(\ \implies r^2\partial_uQ_c=0 \)\\
\\
\begin{equation*}
  \partial_uC= \partial_uQ_c=0 \tag{IX} \label{sub:9}
\end{equation*}
\\
\\
from \eqref{eqn10}\\
\\
\begin{equation*}
  \partial_\phi Q_c +\sin^2\theta\partial_\theta Q_d=0   \tag{X} \label{sub:10}
\end{equation*}
\\
\\
From \eqref{eqn4}\\
\\
\(\ B\sin\theta+C(r\cos\theta)+r\sin\theta\partial_\phi D=\frac{rk\sin\theta} {2} \)\\
\\
\(\ \implies r(k-\partial_u A) \sin\theta+rQ_c\cos\theta +r\sin\theta\partial_\phi D=\frac{rk\sin\theta}{2} \)\\
\\
\(\ \implies \frac{k\sin\theta} {2}-\sin\theta\partial_u A+Q_c\cos\theta+sin\theta\partial_\phi Q_d=0 \)\\
\\
taking $\partial_\phi$ on both sides\\
\\
\(\ \cos\theta\partial_\phi Q_c+\sin\theta\partial_\phi^2 Q_d=0 \)\\
\\
\begin{equation*}  
\cos\theta\partial_\phi Q_c+\sin\theta\partial_\phi^2 Q_d=0 \tag{XI} \label{sub:11}
\end{equation*} 
\\
\\
\\
\\
Furthermore , by using equation \eqref{eqn4} and substituting the value of B from equation \eqref{sub:2} and \eqref{sub:5},\eqref{sub:6}, we obtain the following\\
\\
\(\ C\cos\theta=Q_c\cos\theta=(\partial_uA-\partial_\phi Q_d-\frac{\kappa}{2})\sin\theta \)\\
\\
Differentiate both sides by $\partial_u$ \\
\\
\(\ \cos\theta\partial_u C=(\partial_u^2A-\partial_{\phi u}Q_d)\sin\theta \)\\
\\
From result (\eqref{sub:9}) and (\eqref{sub:8}) we obtain\\
\\
\begin{equation*}
   \partial_u^2A=0 \tag{XII} \label{sub:12}
\end{equation*}
\\
\\
\\
\\
substitute \eqref{sub:2} , \eqref{sub:5} ,\eqref{sub:6} in \eqref{eqn1}
\\
\\
\(\ \frac{1}{r}\frac{3m-r}{r}\partial_uA +\frac{m'}{r^2}A =\frac{1}{r^2}(2m\kappa-\kappa \frac{r} {2}-\frac {2m'}{r^2}) \)\\
\\
\(\ \implies (3m-r)\partial_uA+m'A=(2m-\frac{r}{2})\kappa-\frac{2m'}{r} \)\\
\\
\begin{equation*}    
\partial_u{( (3m-r)A) } = (2m-\frac{r}{2})\kappa - \frac{2m'}{r} \label{sub:15} \tag{XIV}
\end{equation*} \\
\\
Integrate on both sides \\
\\
\(\ \implies A=\frac{1}{3m-r}\Bigl(\int\bigl((2m-\frac{r}{2})\kappa \bigl) du -\frac{2m}{r}+\chi_1 \Bigl) \)\\
\\
Since $\partial_r A=0$, hence Left hand side of the equation is independent of r  \\
\\
Differentiate both sides by $\partial_r$\\
\\
\(\ 0=\frac{1} {(3m-r)^2}\Bigl(\int(4m-r)\frac{\kappa}{2}du +\chi_1 \Bigl) +\frac{1} {3m-r}\bigl(\frac{-\kappa u} {2} +\partial_r\chi_2 \bigl) \)\\
\\
Note here that $\chi_2=\chi_1-\frac{2m}{r}$. Upon grouping terms we obtain \\
\\
\(\ 0=\frac{A{}}{3m-r}+\frac{1}{3m-r}\Bigl(-\frac{\kappa u}{2}+\partial_r\chi_2\Bigl) \)\\
\\
\begin{equation*}
   \implies A=\frac{\kappa u}{2}-\partial_r\chi_2 \tag{XV} \label{sub:14} \\
\end{equation*}
upon taking $\partial_{ur}$ on both sides, $\partial_{ur^2}\chi_2=0 $ implies $m'=0$
\\
\\
Now, substitute \eqref{sub:14} and $m'=0$ in the \eqref{sub:15} and differentiate on both sides by $\partial_u$\\
\\
$$\implies1=\frac{4m-r} {3m-r} $$
$$\implies 4m-r=3m-r \iff m=0 $$
proceeding note that \\
\\
\{from equations \eqref{eqn2},\eqref{sub:6},\eqref{sub:7} \} \\
\\
\(\ \bigl(\partial_r, \partial_\theta, \partial_\phi \bigl)A =0\)\\
 \\
\(\ \implies\bigl(\partial_r^2,\partial_{\theta r}, \partial_{\phi r} \bigl)\chi_2=0\)\\
\\
\(\ \bigl(\partial_r,\partial_\theta,\partial_\phi)\partial_r\chi_2=(0,0,0)\)\\
\\
$\iff \partial_r\chi_2=\Psi $\\
\\
Where $\Psi$ is some arbitrary constant. Observe that \(\partial_{ur} \chi_2=0\) (one may verify this by writing $\chi_2$ in its expanded form).\\ 
$$ \implies\boxed{  A=\frac{ \kappa u} {2} +\Psi} $$
\\
\\
Substituting the value of A and the value of $Q_b$ in eqn\eqref{sub:2}ref we obtain the following result:-
\(\
\implies B =r(\kappa-\partial_uA) = r(\kappa-\frac{\kappa}{2}) =(\frac{\kappa r}{2}) \) \\
\\
$$ \implies \boxed{ B=\frac{\kappa r}{2}}$$
\\
\\
From equation\{\eqref{sub:10},\eqref{sub:11}\} we obtain\\
\(\ \partial_\phi^2Q_d-\sin\theta\cos\theta*\partial_\theta Q_d=0\)\\
This is a partial differential equation. We shall use seperation of varaibles of this to obtain the value of \(Q_d\) .
And subsequently substitute this value in equation \{\eqref{sub:4} ,\eqref{sub:7} ,\eqref{eqn4}\} to obtain the values of C, D. \\
The result [calculations are shown in the appendix] is :-
$$ C=-\sqrt{\Gamma}\big(\psi_1\exp(\phi\sqrt{\Gamma} )-\psi_2\exp(-\phi\sqrt{\Gamma})\bigl)\tan^{\Gamma+1}\theta $$
$$ D=\bigl(\psi_1\exp(\phi\sqrt{\Gamma} )+\psi_2\exp(-\phi\sqrt{\Gamma})\bigl) \tan^{\Gamma}\theta $$

Substitute the above values of C and B in \eqref{eqn3}
\begin{align*}
    rB+r^2\partial_\theta C=\frac{\kappa r^2}{2} \\
   \implies  \frac{\kappa r^2}{2} + r^2\partial_\theta C=\frac{\kappa r^2}{2} \\
   \implies \sqrt{\Gamma}(\Gamma+1)\big(\psi_1\exp(\phi\sqrt{\Gamma} )-\psi_2\exp(-\phi\sqrt{\Gamma})\bigl)\tan^\Gamma\theta=0
\end{align*}
Note that generically the solution of C and D obtained are of the form 
\begin{align*}
    \boxed{C=\sum_\Gamma(-\sqrt{\Gamma}\big(\psi_1\exp(\phi\sqrt{\Gamma} )-\psi_2\exp(-\phi\sqrt{\Gamma})\bigl)\tan^{\Gamma+1}\theta)} 
    \\
    \boxed{ D=\sum_{\Gamma}\bigl(\psi_1\exp(\phi\sqrt{\Gamma} )+\psi_2\exp(-\phi\sqrt{\Gamma})\bigl) \tan^{\Gamma}\theta}
\end{align*}
However due to the linearity of the $\partial_\theta$ operator and the fact that the functions forming C are linearly Independent , it is sufficient to show that it holds for any arbitrary $\Gamma$ term involved.
Also note that the solution set $\psi_1,\psi_2=0$ is a subset of  $ \Gamma=0 $ and the solution set $\Gamma=0$ yields $C=0$ and $D=\psi_1+\psi_2=\psi_3$.
Hence we get the general value of the coefficients of the vector field as
\begin{align*}
 A&=\frac{\kappa  u} {2} +\Psi \tag{R1} \label{r1}
 \\
 B&=\frac{\kappa r} {2} \tag{R2} \label{r2}
 \\
 C&=0 \tag{R3} \label{r3}
 \\ 
 D&=\psi_3\tag{R4} \label{r4}
\end{align*}
Where $\kappa=2\beta-(p+\frac{1}{2})$ ;  \{ $\psi_3,\Psi \} \in  \R$ 

\vspace{3cm}
\newpage
\section{Classification of Geometric Flows on Vaidya Spacetime}
\vspace{0.25cm}
We shall now state the definitions we are going to use to define what a geometric flow is on a Manifold based on what is states in the paper \cite{lauret2015geometricflowssolitonshomogeneous}
\begin{definition}
   Given a time evolving metric on spacetime $(M,g_{\mu\nu})$ . A Homogeneous Geometric flow is defined via the following partial differential equation . 
   \begin{equation*}
       \partial_t g_{\mu\nu} = F(g_{\mu\nu}, \nabla g_{\mu\nu}, R_{\mu\nu}, \ldots)
   \end{equation*}
   where F is a tensor valued operator which produces a tensor of the same kind (in our case its a Symmetric rank (0,2) tensor ) which satisfies two properties.\\
   (i) Given a family of time dependent diffeomorphisms  $l_{t}$ on the manifold,F is invariant under the pullback ($l^*$)
     \begin{equation*}
         l^*(F(g_{\mu\nu}, \nabla g_{\mu\nu}, R_{\mu\nu}, \ldots))=F(l^*g_{\mu\nu}, l^*\nabla g_{\mu\nu}, l^*R_{\mu\nu}, \ldots)
     \end{equation*}\\
   (ii) $\exists m \in \R$ such that $ \forall c\in \R$ 
   \begin{equation*}
   F(cg_{\mu\nu}, \nabla g_{\mu\nu}, R_{\mu\nu}, \ldots) = c^m F(g_{\mu\nu}, \nabla g_{\mu\nu}, R_{\mu\nu}, \ldots)
   \end{equation*}
   A geometric Flow is defined just by the verification of the first property.The inclusion of the second property is what makes it a Homogeneous flow.
\end{definition}
\vspace{0.5cm}
Similarly we state the definition of soliton as taken in the paper \cite{lauret2015geometricflowssolitonshomogeneous}
\begin{definition}
     A solution is defined as self-similar if given $c(t_0)\in \R
     $\begin{equation*}
g (t) = c_2 (t) l^* _t (g (0)) \end{equation*}
\end{definition}

\vspace{0.5cm}

 As defined in \cite{chow2004ricci} and \cite{catino2017ricci}. We follow them in classifying the flow as expanding, steady or contracting on the basis on this c(t) scaling factor. more specifically 
 \\
 \begin{align*}
      c'(0)>0 \implies \textit{flow is expanding}  \\
      c'(0)=0 \implies \textit{flow is steady} \\
      c'(0)<0 \implies \textit{flow is shrinking}
  \end{align*}
\\
As stated in \cite{alhouiti2024geometric} they classified the flow as shrinking , expanding and steady on the basis of the factor $\beta$ .This is because $\beta$ corresponds to $c'(0)$, i.e \\
\begin{align*}
      \beta>0 \implies \textbf{flow is expanding}  \\
      \beta=0 \implies \textbf{flow is steady} \\
      \beta<0 \implies \textbf{flow is shrinking}
  \end{align*}
\\
As one may note from observing from the above $c'(0)$ that we are only analyzing the change in the metric at time $t=0^+$. The core reason for doing such a thing is that this methodology serves as a powerful litmus test whether our solution could exist or not.This is based on the principle of linear stability. The basic intuitive summary of it is that we are analyzing the change in the metric for infinitesimally small perturbation to the system. should the system be initially unstable, due to the nature of the problem, we can safely discard the solution. This is because a solution that is linearly unstable will definitely be unstable in a larger timestep as well.The prospect of analysis for the soliton beyond the initial timestep falls under the domain of non linear stability analysis. Proving nonlinear stability is a monumental challenge and is active in the forefront of research in mathematical relativity, as deduced by the enormous efforts undertaken to prove the stability of the Kerr black hole family.
For the sources of the above information and for further reading, refer to the following sources \cite{regge1957stability},\cite{klainerman2022brief},\cite{chandrasekhar1998mathematical}, \cite{green2020teukolsky}.

\vspace{0.5cm}
\section{The Conformal Gradient Ricci Bourguignon Soliton with respect to Vaidya Metric} 
\vspace{0.25cm}
\begin{definition} \label{def8}
    Given a Manifold M and $X\in\Gamma(TM) $,  $f\in C^{\infty}M $ is defined as the scalar potential of X iff $\nabla f=X $
\end{definition}

\begin{definition} \label{def9}
    Given manifold M, \(f\in C^{\infty}M \) ; \(\nabla f =g^{kj}\frac{\partial f} {\partial x_k}\partial_j \) is defined as the gradient of the scalar field. 
\end{definition}
\vspace{0.25cm}

Applying the definition\{\eqref{def8},\eqref{def9}\} for Vaidya spacetime we obtain the expression for the gradient $f \in C^{\infty}M  $ as the following :-  \\
\\
$$\nabla f=-(\partial_rf)\partial_u - \biggl(\partial_uf+\Bigl(\frac{2m-r} {r}\Bigl) \partial_rf \biggl)\partial_r +\frac{1}{r^2}\bigl(\partial_\theta f) \partial_\theta +\frac{1}{r^2\sin^2\theta}(\partial_\phi f)\partial_\phi $$ \\

$$=A^\mu\frac{\partial}{\partial x_\mu}=A\partial_u+B\partial_r +C\partial_\theta+D\partial_\phi$$
\\
\\
Upon equating the coefficients, we get :-\\
\begin{align*}
    -\partial_rf=A=\frac{\kappa u} {2} + \Psi \tag{4.1}\label{4.1} \\ 
    \partial_uf+\Big(\frac{2m-r}{r}\Big)\partial_rf=-B=-\frac{\kappa r}{2} \tag{4.2}\label{4.2}\\
    \frac{\partial_\theta f}{r^2} =C \tag{4.3}\label{4.3}\\
    \frac{\partial_\phi f} {r^2\sin^2\theta}=D \tag{4.4}\label{4.4}
\end{align*}
\\
\(\ -\partial_rf=A=\frac{\kappa u} {2} + \Psi \)\\
\\
\begin{equation*}
\iff \boxed{f=-\frac{\kappa ur} {2}-\Psi r  +\Pi^*_a(u,\theta,\phi) }  \tag{G1}\label{sr1}
\end{equation*}
\\
Similarly doing the same for B 
\\
\(\ \partial_uf+\Big(\frac{2m-r}{r}\Big)\partial_rf=-B=-\frac{\kappa r}{2}\) \\
\\
We know that m=0.we further substitute equation \eqref{4.1} in \eqref{4.2} \\
\(\ \partial_uf=-(A+B)=-(\frac{ku} {2} + \frac{kr} {2} +\Psi) \)
\\
\begin{equation*}
\iff \boxed{f=-\frac{kur} {2}-\frac{ku^2} {4} - \Psi u +\Pi^*_b(r, \theta, \phi)} \tag{G2}\label{sr2}
\end{equation*}
\\
\\

Similarly, \(\frac{\partial_\theta f}{r^2} =C\) and\(\frac{\partial_\phi f} {r^2\sin^2\theta}=D \)    .  \\
\begin{align*}
    C=0 =\partial_\theta f\\ D=\psi_3\\
    \implies \partial_\phi f=\psi_3r^2\sin^2\theta\\
    \implies \partial_{r\phi}f=0=2\psi_3r\sin^2\theta\\
    \implies \psi_3=0
\end{align*}
Therefore , for scalar potential to exist $\psi_3=0 $ as a condition as shown above.
Hence $\partial_\phi f=0$ , $\partial_\theta f=0$ . Therefore apply the value of $f$ from \eqref{sr1},\eqref{sr2}
we obtain the following.
\begin{align*}
    \partial_\theta f=0=\partial_\theta\Pi^*_a(u,\theta,\phi) \implies \Pi^*_a(u,\phi)\\
    \partial_\phi f=0=\partial_\phi\Pi^*_a(u,\phi)\implies\boxed{\Pi^*_a=\Pi^*_a(u)} \\
    \textbf{similarly}\\
    \boxed{\Pi^*_b=\Pi^*_b(r)}
\end{align*}

\vspace{0.5cm}
from eqns(\eqref{sr1} ,\eqref{sr2} ) and differentiate both of them by $\partial_r$ and equate the $\partial_r f$ \\
\\
\(\iff f=-\frac{\kappa ur} {2}-\Psi r+\Pi^*_a \hspace{0.7cm}   and \hspace{00.7cm}    f=-\frac{\kappa ur}{2}-\frac{\kappa u^2}{4} -\Psi u + \Pi^*_b\) \\
\\
\(\implies \partial_rf=-\frac{\kappa u}{2}-\Psi=-\frac{\kappa u} {2}+\partial_r\Pi^*_b\)\\
\\
\(\implies \partial_r\Pi^*_b=-\Psi \iff \Pi^*_b=-\Psi r +\Pi^{**}_b(\theta, \phi) \)\\
\\
Similarly we $\partial_u$ and equate \(\partial_uf\) from eqn(\eqref{sr1} ,\eqref{sr2}) \\
\\
\(\implies \partial_uf=-\frac{\kappa r}{2}+\partial_u\Pi^*_a=-\frac{\kappa r}{2}-\frac{\kappa u} {2}-\Psi \)
\\
\(\iff \partial_u\Pi^*_a=-\frac{\kappa u} {2}-\Psi \implies \Pi^*_a=-\frac{\kappa u^2}{4}-\Psi u-\Pi^{**}_a(\theta,\phi)\) 
\\
We hence get the result 
\(\implies f=-\frac{\kappa ur}{2}-\frac{\kappa u^2}{4}-\Psi(r+u)+\Pi^{**}_a(\theta,\phi)\)\\
\\
Applying this equation to ,Since $\partial_\theta f=0$ and $\partial_\phi f=0$ we obtain
\(\Pi^{**}_a=\Psi_2 \) where $\Psi_2 \in \R $
\\
\\
Note : The Vector field has to be Irrotational depending upon the boundary conditions else such a scalar potential cannot exist.There for the final form of $f\in C^{\infty}M $ as a scalar potential function of conformal ricci bourguignon soliton is \\
\begin{equation*}
\boxed{ f=-\frac{\kappa u}{2} (r-\frac{u}{2}) - \Psi(r+u)+\Psi_2} 
 \tag{R5}\label{r5}
\end{equation*}

\newpage

\section{Summary and Results} 
\vspace{0.1cm}
The general value of the coefficients of the vector field as
\begin{align*}
 A&=\frac{\kappa  u} {2} +\Psi 
 \\
 B&=\frac{\kappa r} {2} 
 \\
 C&=0 
 \\ 
 D&=\psi_3
\end{align*}
Where $\kappa=2\beta-(p+\frac{1}{2})$ ;  \{ $\psi_3,\Psi \} \in  \R$
\\
\\
\\
For the vector field to have a scalar potential, $\psi_3=0$ .  the resulting potential is
$$f=-\frac{\kappa u}{2} (r-\frac{u}{2}) - \Psi(r+u)+\Psi_2$$ 
Where $ \{  \psi_2, \Psi\}\in \R$\\
\\
\\
Furthermore, this Soliton is classified as steady, shrinking, expanding on the basis of the constant $\beta$
\begin{align*}
      \beta>0 \implies \textbf{flow is expanding}  \\
      \beta=0 \implies \textbf{flow is steady} \\
      \beta<0 \implies \textbf{flow is shrinking}
  \end{align*}
\\

We have derived the general solution for a conformal Ricci Bourguignon soliton in Vaidya Spacetime and showed that for the soliton to exist the metric gets reduced to the Schwartzschild metric with $m=0$. This solution serves as a starting point for exploring various physical phenomena and situations by applying a variety of boundary conditions , and can be used to study the soliton's behavior under perturbations.
\newpage
\section{Appendix}
\vspace{0.25cm}
[A]\textbf{Proof involving m=0 $\implies \partial_\theta A=0 \text{ and }\partial_\phi A=0$} \\
We had earlier assumed at (\ref{sub:6}) that $\partial_\theta A=0 \text{ and } \partial_\phi A=0\implies m=0$.Now we shall prove that $m=0\implies \partial_\theta A=0 \text{ and }\partial_\phi A=0$ 
\\
\\
Suppose m=0, then the differential equations \eqref{eqn1}\eqref{eqn6}\eqref{eqn7} are reduced to the following.
\begin{align*}
 \partial_u(A+B)=\frac{\kappa}{2}\\
 \partial_\theta(A+B)=r^2\partial_uC \\
 \partial_\phi(A+B)=r^2\sin^2\theta\partial_uD
\end{align*}
from \eqref{sub:2} and \eqref{sub:5} \\
\(B=(\kappa-\partial_uA)r+Q_b\)\\
\(Q_b=\partial_\theta^2A\)\\
from \eqref{eqn6}\\
\(\partial_\theta(A+B)=r^2\partial_uC\)\\
\(\implies\partial_\theta(A+(\kappa-\partial_uA)r+\partial_\theta^2A)=r^2\partial_u(-\frac{\partial_\theta A}{r}+Q_c)\)\\
differentiate on both sides by $\partial_r$\\
\(\implies\partial_uQ_c=0 \) and 
\(    \partial_\theta A+\partial_\theta^3A=0 \)
upon performing a similar procedure to equation \eqref{eqn7} we get\\
\(\partial_\phi(A+\partial_u^2A)=0\)\\
From equation \eqref{eqn2} we further obtain the following result. \\
\(\partial_r(A+\partial_\theta^2A)=0\) \\
Therefore, from above , \(A+\partial_\theta^2A=f_3(u)\)
from \eqref{eqn1}\\
\(\partial_u(A+B)=\frac{\kappa}{2}\)\\
\(\implies \partial_uA-r\partial_u^2A + \partial_{u\theta^2}A=\frac{\kappa}{2}\)\\
differentiate on both sides by $\partial_r$ and substituting the result back into the equation, we get:-\\
\(\implies \partial_u^2A=0 \text{ and } \partial_u(A+\partial_\theta^2A)=\frac{\kappa}{2}\)\\
\begin{equation*}  
\implies \boxed{ A+\partial_\theta^2A=\frac{\kappa u}{2} +\psi_3} \tag{A1} \label{a1}
\end{equation*}\\
Substitute the above in \eqref{eqn5} to obtain:-
\(B=\frac{\kappa r}{2}+r\partial_{u\theta^2}A+\partial_\theta^2A\)\\
Upon substituting the above result of B in \eqref{eqn3} , then we substitute the value of C from \eqref{sub:3} we get:-
\(B+r\partial_\theta C=\frac{\kappa r}{2}\)\\
\(\implies\frac{\kappa r}{2}+r\partial_{u\theta^2}A+\partial_\theta^2A-r\frac{\partial_\theta^2A}{r}+r\partial_\theta Q_c=\frac{\kappa r}{2}\)\\
\(\implies \partial_{u\theta^2}A+\partial_\theta C=0\)\\
$\partial_u$ on both sides\{note $\partial_u^2A=0$\}
\(\implies \partial_{\theta u}C=\partial_{\theta^2u^2}A=0 \)\\
\(\implies \frac{\partial_{u\theta^2}A}{r}=0\)\\
\(\partial_{u\theta^2}A=0\)
therefore $\partial_{\theta^2}A=f_2(\theta,\phi)$\\

From here differentiate \eqref{a1} on both sides by $\partial_\theta^2$ to obtain the following:-
\(\partial_\theta^4A+\partial_\theta^2A=0\)\\

\begin{equation*}
    \implies \boxed{\partial_\theta^2A=\Phi_1(\phi)\cos\theta+\Phi_2(\phi)\sin\theta}
\end{equation*}
From equation \eqref{sub:3} , differentiate on both sides by $\partial_r$\\
\(\implies \partial_{\phi}\partial_rC+\sin^2\theta\partial_\theta\partial_rD=0\)\\
\(\implies \frac{\partial_{\phi\theta}A}{r^2}+\sin^2\theta\partial_{\theta}\frac{\partial_\phi A}{r^2\sin^2\theta}=0\)\\

\begin{equation}
    \implies \boxed{\partial_{\phi\theta}A+\partial_\phi A\cot\theta=0} \label{eqn7.1}
\end{equation}
Now take \eqref{eqn4} , divide by $\sin^2\theta$ differentiate it on both sides by $\partial^2_r$ and substitute the values in \eqref{eqn8} , \eqref{eqn9} 
\(\implies -\frac{\partial_\theta A}{r^2}\cot\theta +\frac{\partial_\theta A}{r^2}\cot\theta+2\frac{\partial_\phi^2A}{\sin^2\theta}=0 \)
\(\implies \boxed{\partial^2_\phi A=0}\)\\
\\
\begin{align*}
    \partial_\theta^2A=p_1(\phi)\cos\theta +p_2(\phi)\sin\theta\\
     A=\frac{\kappa u}{2}-\partial_\theta^2A\\
     \implies A=\frac{\kappa u}{2}-\bigl(p_1(\phi)\cos\theta +p_2(\phi)\sin\theta \bigl)\\
     \implies \partial_\phi^2A=0=\cos\theta\frac{d^2p_1}{d\phi^2}+\sin\theta\frac{d^2p_2}{d\phi^2}\\
     \implies \frac{d^2p_1}{d\phi^2} ,\frac{d^2p_2}{d\phi^2}=0\\
     \implies p_1=h_{11}\phi +h_{12} ; p_1=h_{21}\phi +h_{22} \\
     \text{substitute the value of A in eqn [\eqref{eqn7.1}]}\\
     -\cot\theta\partial_{\theta}A=(p_1\sin\theta-p_2\cos\theta)(-\cot\theta) = \partial_{\phi\theta}A=\frac{dp_1}{d\phi}\sin\theta-\frac{dp_2}{d\phi}\cos\theta\\
     \implies \Bigl(\frac{dp_2}{d\phi}-p_1\Bigl)\cos\theta + p_2\frac{\cos^2\theta}{\sin\theta} -\frac{dp_1}{d\phi}\sin\theta =0 \\
     \implies p_1=p_2=\frac{dp_1}{d\phi} =\frac{dp_2}{d\phi}=0
\end{align*}
\begin{align*}
   \boxed{ \partial_\theta A=0} \\ 
    \boxed{\partial_\phi A=0}
\end{align*}
\textbf{Hence we end up with m=0 $\iff \{\partial_\theta A=0 \text{ and } \partial_\phi A=0 \}$}
\newpage
[B]\textbf{Proof involving $\L_Xg(\partial_i,\partial_j)= X^k\partial _k(g_{ij}) + g_{kj}(\partial _i X^k) +g_{ik}(\partial _j X^k)$}\\ \\
refer \cite{Lee2003} edition[2] chapter \{12\} corollary[12.33] 
\begin{align*}
    \L_Xg(\partial_i,\partial_j)=X(g(\partial_i,\partial_j))-g([X,\partial_i],\partial_j)-g(\partial_i,[X,\partial_j]) \\
    = X(g_{ij})-g(X(\partial_i)-\partial_i(X),\partial_j)-g(\partial_i,X(\partial_j)-\partial_j(X))\\
    = X^k\partial_k(g_{ij})-g(X^k\partial_k(\partial_i)-\partial_i(X^l\partial_l),\partial_j)-g(\partial_i,X^n\partial_n(\partial_j)-\partial_j(X^m\partial_m))\\
    =X^k\partial_k(g_{ij})-g(-\partial_i(X^l)\partial_l,\partial_j)-g(\partial_i,-\partial_j(X^m)\partial_m)\\
    =X^k\partial_k(g_{ij})+\partial_i(X^l)g(\partial_l,\partial_j)+\partial_j(X^m)g(\partial_i,\partial_m)\\
    =X^k\partial _k(g_{ij}) + g_{kj}(\partial _i X^k) +g_{ik}(\partial _j X^k)
\end{align*}
Hence we obtain the formula for the lie derivitive of the metric tensor\\
\vspace{1cm}
\\

[C]\textbf{Proof Obtaining The Expression For C and D }
Combine equations (\eqref{sub:10},\eqref{sub:11}) we obtain
\begin{align*}
    \partial_\phi^2Q_d-\sin\theta\cos\theta\partial_\theta Q_d=0\\
    \textbf{set } : Q_d=\Phi(\phi)\Theta(\theta) \\
    \implies \frac{d^2\Phi(\phi)}{d\phi^2}\Theta(\theta)=\sin\theta\cos\theta\frac{d\Theta(\theta)}{d\theta}\\
    \implies \frac{\frac{d^2\Phi}{d\phi^2}}{\Phi}=\frac{\frac{d\Theta}{d\theta}}{\Theta}\sin\theta\cos\theta=\Gamma \\
    \implies \Phi=w_1\exp(\sqrt{\Gamma}\phi)+w_2\exp(-\sqrt{\Gamma}\phi) ; \Theta=w_3\tan^\Gamma\theta\\
    \implies D=Q_d=\bigl(\psi_1\exp(\phi\sqrt{\Gamma} )+\psi_2\exp(-\phi\sqrt{\Gamma})\bigl) \tan^{\Gamma}\theta
\end{align*}
From substituting the value of B in \eqref{eqn4} we can obtain C by \\
$$C=-\partial_\phi D \tan\theta=-\sqrt{\Gamma}\big(\psi_1\exp(\phi\sqrt{\Gamma} )-\psi_2\exp(-\phi\sqrt{\Gamma})\bigl)\tan^{\Gamma+1}\theta$$
\vspace{1cm}

% --- ACKNOWLEDGEMENTS ---
\section*{Acknowledgements}
% This adds it to the Table of Contents

The authors acknowledge the use of Google's Gemini and X's Grok for assistance with language editing of the manuscript.

% --- REFERENCES ---
\newpage
 % This adds it to the Table of Contents

\bibliographystyle{abbrv}
\bibliography{References} % This will import your 'References.bib' file

% ===================================================================
%  END OF DOCUMENT
% ===================================================================
\end{document}